\documentclass{ws-ijgmmp}

\begin{document}

\markboth{Che-Yu Chen}
{On the possible spacetime structures of rotating loop quantum black holes}

%
\catchline{}{}{}{}{}
%

\title{ON THE POSSIBLE SPACETIME STRUCTURES OF ROTATING LOOP QUANTUM BLACK HOLES
}

\author{CHE-YU CHEN}

\address{Institute of Physics, Academia Sinica\\
Taipei, 11529, Taiwan\,\\
\email{b97202056@gmail.com} }

\maketitle


\begin{abstract}
To date, a mathematically consistent construction of effective rotating black hole models in the context of Loop Quantum Gravity (LQG) is still lacking. In this work, we start with the assumption that rotating LQG black hole metrics can be effectively obtained using Newman-Janis Algorithm. Then, based on a few extra fair assumptions on the seed metric functions, we make a conjecture on what a rotating LQG black hole would generically look like. Our general arguments and conclusions can be supported by some known specific examples in the literature.
\end{abstract}

\keywords{Non-singular black hole; Loop quantum gravity}

\section{Introduction}	

The direct detection of gravitational waves emitted from binary black hole mergers, as well as the first image of the supermassive black hole at the center of the M87 galaxy, are two important breakthroughs in modern physics. These two crucial achievements have ushered in a golden era in which exploring strong gravity regimes, such as the vicinity of black holes, can indeed be realized. In particular, it turns out that these extremely compact objects are so far the best candidates for testing General Relativity (GR) or even for testing fundamental quantum theories of gravity, such as Loop Quantum Gravity (LQG). 

LQG is a non-perturbative quantum gravity approach, within which the essence of spacetime is characterized by discretized eigenvalues of a set of geometrical operators. This discrete nature, in particular the non-zero area gap associated with the geometrical operators, plays a key role in removing classical singularities in the theory. Moreover, within the framework of LQG and considering the semi-classical regimes, one can construct various effective (non-rotating) black hole models, with one common feature that the classical singularity is usually replaced by a transition surface that connects a black hole and a white hole regions \cite{Gambini:2013ooa,Corichi:2015xia,Olmedo:2017lvt,Ashtekar:2018lag,Ashtekar:2018cay,Bodendorfer:2019xbp,Bodendorfer:2019cyv,Arruga:2019kyd,Assanioussi:2019twp,BenAchour:2020gon,Gambini:2020nsf,Bodendorfer:2019nvy,Bodendorfer:2019jay,Blanchette:2020kkk,Assanioussi:2020ezr} (see \cite{Bojowald:2020dkb} for a recent review){\footnote{Upon using different quantization approaches, the classical singularities inside a non-rotating black hole could be altered into other interesting scenarios, such as the Nariai spacetime \cite{Boehmer:2007ket} or an Euclidean region \cite{Bojowald:2018xxu}. The formation of the inner horizon is also possible \cite{BenAchour:2018khr,Kelly:2020uwj}.}}. However, the technical difficulties of invoking real-valued Ashtekar-Barbero variables in axisymmetric spacetimes \cite{Frodden:2012en,Gambini:2020fnd} largely hinder the progress of constructing effective LQG models for rotating black holes. In the literature, some attempts along this ling have been made by adopting the Newman-Janis Algorithm (NJA) \cite{Caravelli:2010ff,DeLorenzo:2015taa,Liu:2020ola,Brahma:2020eos}. As a metric-generating method, NJA works quite well in generating Kerr and Kerr-Newman metrics starting with their non-rotating counterparts, called seed metrics \cite{Newman:1965tw}. Although it is challenging at this point to justify the validity of using NJA beyond GR, it may still allow us to construct effective models for rotating black holes, which could capture the key features that LQG black holes are supposed to have.

As we have just mentioned, although there are various ways of constructing effective non-rotating LQG black holes, one common feature is that the classical singularity is replaced by a transition surface, where the radius of the 2-sphere acquires its minimum. Based on this observation, one could naively guess the possible spacetime structure that a rotating LQG black hole could have. Such a conjecture has been made in \cite{Brahma:2020eos}. The authors of \cite{Brahma:2020eos} argued that depending on the relative location of the transition surface and the two horizons{\footnote{One direct consequence when a black hole starts spinning is the appearance of the inner horizon.}}, a rotating LQG black hole could appear as a wormhole, a black hole with one horizon, or a black hole with two horizons. In \cite{Brahma:2020eos}, the rotating counterpart of a particular LQG black hole model \cite{Bodendorfer:2019nvy,Bodendorfer:2019jay} was constructed as a toy model to support the general conjecture of the paper. In addition, possible astrophysical implications and observational consequences were discussed.

In this paper, instead of starting with a specific LQG black hole model, we would like to strengthen the above conjecture on a more general ground. More explicitly, we start with the following assumptions: \textrm{(i)} The resultant metric after performing NJA can effectively describe the rotating counterpart of a LQG black hole. \textrm{(ii)} The non-rotating LQG black holes asymptotically reduce to the Schwarzschild black holes at spatial infinity, and replace the classical singularity with a transition surface. \textrm{(iii)} Quantum corrections start to have substantial effects on the spacetime geometry only at the region sufficiently close to the transition surface. We will show that based on these fair assumptions, the general conjecture made in \cite{Brahma:2020eos} can already be supported. Two specific effective LQG black hole models will also be provided to support our arguments.

The paper is outlined as follows. In sec.~\ref{sec.NJA}, we quickly review how NJA and its revised version work in general to obtain rotating spacetimes from a non-rotating seed metric. Then, in sec.~\ref{sec.assumption}, we put down the minimal set of assumptions on the seed metric functions of a non-rotating LQG black hole. Based on these fair assumptions, we make conjectures on the possible spacetime structures of rotating LQG black holes in sec.~\ref{sec.rotating}. We finally conclude in sec.~\ref{sec.conclusion}.

\section{Revised Newman-Janis Algorithm}\label{sec.NJA}
In this work, we first assume that the rotating LQG black hole metric can be effectively derived using NJA. This assumption allows us to make some quantitative statements on the phenomenology of the model. In this section, we will briefly review NJA \cite{Newman:1965tw} and how its revised version \cite{Azreg-Ainou:2014pra} works.

The NJA starts with a general static and spherically symmetric seed metric:
\begin{equation}
ds^2=-g(y)dt^2+\frac{dy^2}{g(y)}+b(y)^2d\Omega_2^2\,,\label{NJAseed}
\end{equation}
where $g(y)$ and $b(y)$ are metric functions and they are expressed in terms of a radial variable $y$. The seed metric can be recast into the advanced null coordinate system $(u,y,\theta,\phi)$ by defining the following variables
\begin{equation}
u\equiv t-y_*\,,\qquad \frac{dy_*}{dy}\equiv\frac{1}{g(y)}\,.
\end{equation}
In this way, the inverse metric can be expressed using a null tetrad $Z_a^\mu=\left(l^\mu,n^\mu,m^\mu,\bar{m}^\mu\right)$ via
\begin{equation}
g^{\mu\nu}=-l^\mu n^\nu-l^\nu n^\mu+m^\mu\bar{m}^\nu+m^\nu\bar{m}^\mu\,,\label{metrictetrad}
\end{equation}
where $\bar{m}^\mu$ is the complex conjugate of $m^\mu$. The explicit expression of the null tetrad $Z_a^\mu$ is not shown here because it is not very informative (see \cite{Newman:1965tw,Azreg-Ainou:2014pra} for the detailed expression).  

The most important step in NJA is to perform a complex shift on the advanced null coordinates
\begin{equation}
u'=u-ia\cos\theta\,,\qquad y'=y+ia\cos\theta\,,
\end{equation} 
where $a$ will be later regarded as the spin of the spacetime. Note that the angular coordinates $\theta$ and $\phi$ remain unchanged. After the complex shift, the coordinate set becomes $(u',y',\theta,\phi)$. Also, the metric functions can in general be expressed as functions of $y$ and $\theta$ (after dropping the prime for simplicity). Let us denote them as
\begin{equation}
g(y)\rightarrow G(y,\theta)\,,\qquad b(y)^2\rightarrow\Psi(y,\theta)\,,
\end{equation}
for the time being. After the complex shift, the set of null tetrad basis is changed and one can use Eq.~\eqref{metrictetrad} to obtain a new line element in the advanced null coordinates as follows
\begin{align}
ds^2=&-2dudy+2a\sin^2\theta\left(G-1\right)dud\phi-Gdu^2+\Psi d\theta^2+2a\sin^2\theta dyd\phi\nonumber\\
&+\sin^2\theta\left[\Psi+a^2\sin^2\theta\left(2-G\right)\right]d\phi^2\,.\label{NJA1}
\end{align}

At this point, the metric functions $G(y,\theta)$ and $\Psi(y,\theta)$ remain undetermined. They can actually be determined in the last step of NJA, or more precisely, the last step of the revised version of NJA. This step is to rewrite the metric \eqref{NJA1} in the Boyer-Lindquist coordinate system $(t,y,\theta,\varphi)$, in which the $g_{t\varphi}$ component is the only off-diagonal metric component. This can be done by considering the following transformations:
\begin{equation}
du=dt+\lambda_1(y)dy\,,\qquad d\phi=d\varphi+\lambda_2(y)dy\,,\label{coordinatetran}
\end{equation}
where 
\begin{equation}
\lambda_1(y)=-\frac{\Psi(y,\theta)+a^2\sin^2\theta}{G(y,\theta)\Psi(y,\theta)+a^2\sin^2\theta}\,,\quad\lambda_2(y)=-\frac{a}{G(y,\theta)\Psi(y,\theta)+a^2\sin^2\theta}\,.
\end{equation}
As one can clearly see, in order to have well-defined coordinate transformations \eqref{coordinatetran}, $\lambda_1$ and $\lambda_2$ have to be functions of $y$ only. This is in general not possible for arbitrary $G(y,\theta)$ and $\Psi(y,\theta)$. Even if one considers the standard complexification procedure in the original NJA, the coordinate transformations \eqref{coordinatetran} are still not guaranteed to be well-defined. However, in the revised NJA \cite{Azreg-Ainou:2014pra}, the metric functions $G(y,\theta)$ and $\Psi(y,\theta)$ are chosen exquisitely such that the transformations \eqref{coordinatetran} are ensured to be well-defined. More explicitly, the metric functions are chosen to be related to those of the seed metric as follows
\begin{equation}
G(y,\theta)=\frac{g(y)b(y)^2+a^2\cos^2\theta}{\Psi(y,\theta)}\,,\qquad \Psi(y,\theta)=b(y)^2+a^2\cos^2\theta\,.
\end{equation}
Then, the functions $\lambda_1$ and $\lambda_2$ are ensured to be functions of $y$ and the transformations \eqref{coordinatetran} are well-defined.

After imposing the coordinate transformations, the final expression of the rotating spacetime, in the Boyer-Lindquist coordinates, can be written as
\begin{equation}
ds^2=-\left(1-\frac{2Mb}{\rho^2}\right)dt^2-\frac{4aMb\sin^2\theta}{\rho^2}dtd\varphi+\rho^2d\theta^2+\frac{\rho^2dy^2}{\Delta}+\frac{\Sigma\sin^2\theta}{\rho^2}d\varphi^2\,,\label{NJAmetricfinal}
\end{equation}
where
\begin{align}
\rho^2=b(y)^2+a^2\cos^2\theta\,,\quad M=M(y)\equiv b(y)\left(1-g(y)\right)/2\,,\nonumber\\
\Delta=\Delta(y)\equiv g(y)b(y)^2+a^2\,,\quad \Sigma=\left(a^2+b(y)^2\right)^2-a^2\Delta\sin^2\theta\,.
\end{align}

\section{Assumptions on the seed metric}\label{sec.assumption}
Having written down the general expression of the rotating metric \eqref{NJAmetricfinal}, one can observe that the rotating metric can actually be expressed in terms of the seed metric functions. Therefore, any assumptions or restrictions on the seed metric \eqref{NJAseed} would have direct consequences on the rotating metric. In this section, we will outline the assumptions on the seed metric functions $g(y)$ and $b(y)$ before we further discuss the possible spacetime structures of the rotating spacetime.

We assume that the seed metric for non-rotating LQG black holes has the following properties
\begin{arabiclist}
\item The seed metric has a non-degenerated event horizon at $y=y_h\ne0$, such that $g(y_h)=0$, $g'(y_h)>0$, and $b(y_h)\ne0$, where the prime denotes the derivative with respect to $y$.
\item The quantum effects replace the classical singularity by a spacelike transition surface inside the event horizon. without loss of generality, we assume that the transition surface is located at $y=0$ at which the radius of the 2-sphere has a minimum value, i.e., $b(0)=b_0>0$ and $b'(0)=0$. Also, since the transition surface is spacelike, we require $g(0)<0$. The transition surface at $y=0$ connects a black hole region on one side and a white hole region on the other side. The regions of a positive (negative) $y$ correspond to the black (white) hole regions. To be more explicit, we may assume that the metric functions near the transition surface can be expanded as follows:
\begin{equation}
g(y)=-|g_0|+g_2y^2+\mathcal{O}(y^3)\,,\quad b(y)=b_0+b_2y^2+\mathcal{O}(y^3)\,.\label{taylory0}
\end{equation}
Note that upon requiring the spacetimes to be symmetric in the black hole and white hole regions at least near the transition surface, one can have $g'(0)=0$.
\item The seed metric asymptotically reduces to the Schwarzschild one, namely, when $|y|\rightarrow\infty$, we have
\begin{equation}
b(y)\rightarrow |y|\,,\qquad g(y)\rightarrow 1-\frac{2M_B}{b(y)}\rightarrow1-\frac{2M_B}{|y|}\,,
\end{equation}
where $M_B$ is the Arnowitt-Deser-Misner mass of the black hole. This assumption ensures that the rotating counterpart \eqref{NJAmetricfinal} asymptotically reduces to Kerr metric when $|y|\rightarrow\infty$. 
\item Quantum effects are sizable only in the vicinity of the transition surface.
\end{arabiclist}

The last assumption can be put down in a more mathematical manner. Essentially, if we collect all the quantum effects and regard their associated geometric corrections as an \textit{effective} matter content, the LQG black hole model can be formally governed by the Einstein equation with an effective energy-momentum tensor on the right-hand side:
\begin{equation}
G_{\mu\nu}=T^{\textrm{eff}}_{\mu\nu}\,,
\end{equation}
where $G_{\mu\nu}$ is the Einstein tensor and we have used the convention $8\pi G_N=1$. Naively, the effective energy-momentum tensor associated with quantum effects would violate energy conditions such that it can provide sort of repulsive forces to prevent the gravitational collapse. In this regard, the last assumption can be translated into the requirement that \textit{the effective energy-momentum tensor satisfies the strong energy condition inside the event horizon, except for the region very close to the transition surface.}

The fulfillment of energy conditions can actually restrict the behavior of the metric functions significantly. First, according to the results in \cite{Yang:2021civ}, if the strong energy condition is satisfied inside a static black hole, there is at most one non-degenerated inner horizon inside every connected branch of a black hole event horizon. Since in our case, the transition surface inside the event horizon is spacelike, therefore, there cannot exist any inner horizon within the region where strong energy condition is satisfied. As long as the strong energy condition can possibly be violated only very close to the transition surface, the spacetime has only one horizon at $y=y_h$ on one side of the transition surface, i.e., $g(y)$ has a single root at $y=y_h$ in the region $y\ge0$.

Indeed, the fulfillment of the strong energy condition inside the event horizon of the seed metric \eqref{NJAseed} implies
\begin{equation}
I_1(y)\equiv\frac{2b'(y)g'(y)}{b(y)}+g''(y)+4g(y)\frac{b''(y)}{b(y)}\ge0\,,\qquad I_2(y)\equiv g(y)\frac{b''(y)}{b(y)}\ge0\,.\label{sec}
\end{equation} 
To proceed, we define the function 
\begin{equation}
F(y)\equiv g(y)b(y)^2\,,
\end{equation}
and consider the following combination of its derivatives
\begin{equation}
\frac{1}{b(y)^2}\left[F''(y)-\frac{2b'(y)}{b(y)}F'(y)\right]=I_1(y)-2g(y)\frac{b'(y)^2}{b(y)^2}-2I_2(y)\,.\label{seccheck}
\end{equation}
According to the first inequality of \eqref{sec} and due to the fact that $g(y)<0$ inside the horizon, the combination of the first two terms on the right-hand side of Eq.~\eqref{seccheck} is non-negative. In addition, we further assume that in the region where strong energy condition is satisfied, the metric function $b(y)$ can already be well approximated by $|y|$ such that the contribution from $I_2(y)$ is negligible. Based on these arguments and assumptions, the left-hand side of Eq.~\eqref{seccheck} is non-negative, meaning that the function $F(y)$ cannot have local maxima inside the horizon, except for the region very close to the transition surface. For simplicity, we will directly assume that the local maximum of the function $F(y)$, if it has any, can only appear at the transition surface. 

In the next section, we will show that, if the rotating counterpart of a LQG black hole can be described effectively by the metric \eqref{NJAmetricfinal}, its spacetime structure can already be substantially determined based on the above assumptions.

\section{Spacetime structure of a rotating LQG black hole}\label{sec.rotating}
According to the rotating metric \eqref{NJAmetricfinal}, the event horizon of the metric is given by 
\begin{equation}
\Delta(y)\equiv F(y)+a^2=0\,.
\end{equation}
Note that when $a=0$, the horizon is located at $y=y_h$ because $g(y_h)=0$. Therefore, the question about how many event horizons could be there reduces to the following mathematical question: \textit{How many roots of the equation $F(y)+a^2=0$ could be there in the region $y>0$?}

The first observation is that once a non-zero spin $a$ is included, the function $F(y)+a^2$ becomes positive at $y=y_h$. Outside $y_h$, the metric functions $g(y)$ and $b(y)$ gradually reduce to their Schwarzschild counterparts, based on the assumption 3. Therefore, the number of horizons is determined by the number of roots of $F(y)+a^2$ within the region $0<y<y_h$, and this turns out to strongly depend on the sign of $F(0)+a^2$.

Let us split the discussion into the value of $F(0)+a^2$ being negative, zero, and positive separately:
\begin{itemize}
\item $F(0)+a^2<0$:\\
In this case, it is apparent that there must be at least one root in the region $0<y<y_h$ because $F(y_h)+a^2>0$. In addition, if we adopt the assumption that the function $F(y)$ can possibly have a local maximum only at $y=0$ (assumption 4), then $F(y)+a^2$ can only have a single root between $y=0$ and $y=y_h$. Therefore, the spacetime contains one event horizon, behind which hides a spacelike transition surface.
\item $F(0)+a^2=0$:\\
In this case, the transition surface itself becomes an event horizon of the spacetime. The detailed spacetime structure then depends on whether the function $F(y)+a^2$ at $y=0$ is a local maximum or a local minimum. Mathematically, we can use Eqs.~\eqref{taylory0} to obtain the expansion of the function $F(y)+a^2$ near the transition surface:
\begin{align}
F(y)+a^2&=F(0)+a^2+\left(b_0g_2-|g_0|b_2\right)y^2+\mathcal{O}(y^3)\nonumber\\
&=\left(b_0g_2-|g_0|b_2\right)y^2+\mathcal{O}(y^3)\,.\label{quadraticcoeff}
\end{align}
If the quadratic coefficient is positive (negative), the function $F(y)+a^2$ has a local minimum (maximum) at $y=0$.
\begin{itemize}
\item Local minimum at $y=0$ ($b_0g_2-|g_0|b_2>0$):\\
In this case, the function $F(y)+a^2$ monotonically increases with respect to $y$ in the region $0<y<y_h$. Therefore, the transition surface itself is the only event horizon in $y\ge0$ and it is a null surface.
\item Local maximum at $y=0$ ($b_0g_2-|g_0|b_2<0$):\\
In this case, when $y$ increases, the function $F(y)+a^2$ would first decrease and then increase up to $y=y_h$. Therefore, in addition to the transition surface itself, there is another event horizon outside the transition surface. We then have a non-singular black hole with two horizons, with the transition surface itself being the inner horizon.
\item Undetermined case ($b_0g_2-|g_0|b_2=0$):\\
It is indeed possible that the coefficients in the expansions \eqref{taylory0} are so fine-tuned that the quadratic coefficient in Eq.~\eqref{quadraticcoeff} vanishes. In this case, whether $F(y)+a^2$ at $y=0$ is really a local maximum or minimum depends on the sign of the higher-order coefficients in the expansion. However, the following argument is still valid: If the function $F(y)+a^2$ has a local minimum (maximum) at $y=0$, there is one (two) event horizon(s) in $y\ge0$.
\end{itemize}
\item $F(0)+a^2>0$:\\
In this case, the function $F(y)+a^2$ is positive both at $y=0$ and $y=y_h$. Similar to the case for $F(0)+a^2=0$, the number of horizons also depends on whether the function $F(y)+a^2$ has a local maximum or a local minimum at $y=0$.
\begin{itemize}
\item Local minimum at $y=0$:\\
This case includes both the possibilities that $b_0g_2-|g_0|b_2>0$ as well as the fine-tuned case where $b_0g_2-|g_0|b_2=0$ while higher-order coefficients imply a local minimum at $y=0$. In this case, there is no root for $F(y)+a^2=0$ between $y=0$ and $y=y_h$. Therefore, there is no event horizon in the spacetime. The transition surface is naked and is a timelike surface, giving rise to a wormhole geometry.
\item Local maximum at $y=0$:\\
This case includes both the possibilities that $b_0g_2-|g_0|b_2<0$ as well as the fine-tuned case where $b_0g_2-|g_0|b_2=0$ while higher-order coefficients imply a local maximum at $y=0$. In this case, the number of horizons depends on whether the local minimum of $F(y)+a^2$ between $y=0$ and $y=y_h$, say, $y=y_m$, is positive, zero, or negative.
\begin{itemize}
\item If $F(y_m)+a^2>0$, there is no root within $0<y<y_h$. Therefore, the spacetime has no event horizon and becomes a wormhole with a timelike transition surface.
\item If $F(y_m)+a^2=0$, the local minimum is zero. In this case, there is a degenerated horizon at $y=y_m$. The transition surface is inside the degenerated horizon and is timelike. 
\item If $F(y_m)+a^2<0$, there are two event horizons, one exterior horizon at $y_m<y<y_h$ and the other inner horizon at $0<y<y_m$. The spacetime is a non-singular black hole with two horizons. The transition surface is timelike and hidden inside the inner horizon.
\end{itemize}
\end{itemize}
\end{itemize}

\begin{figure}[t]
\centerline{\psfig{file=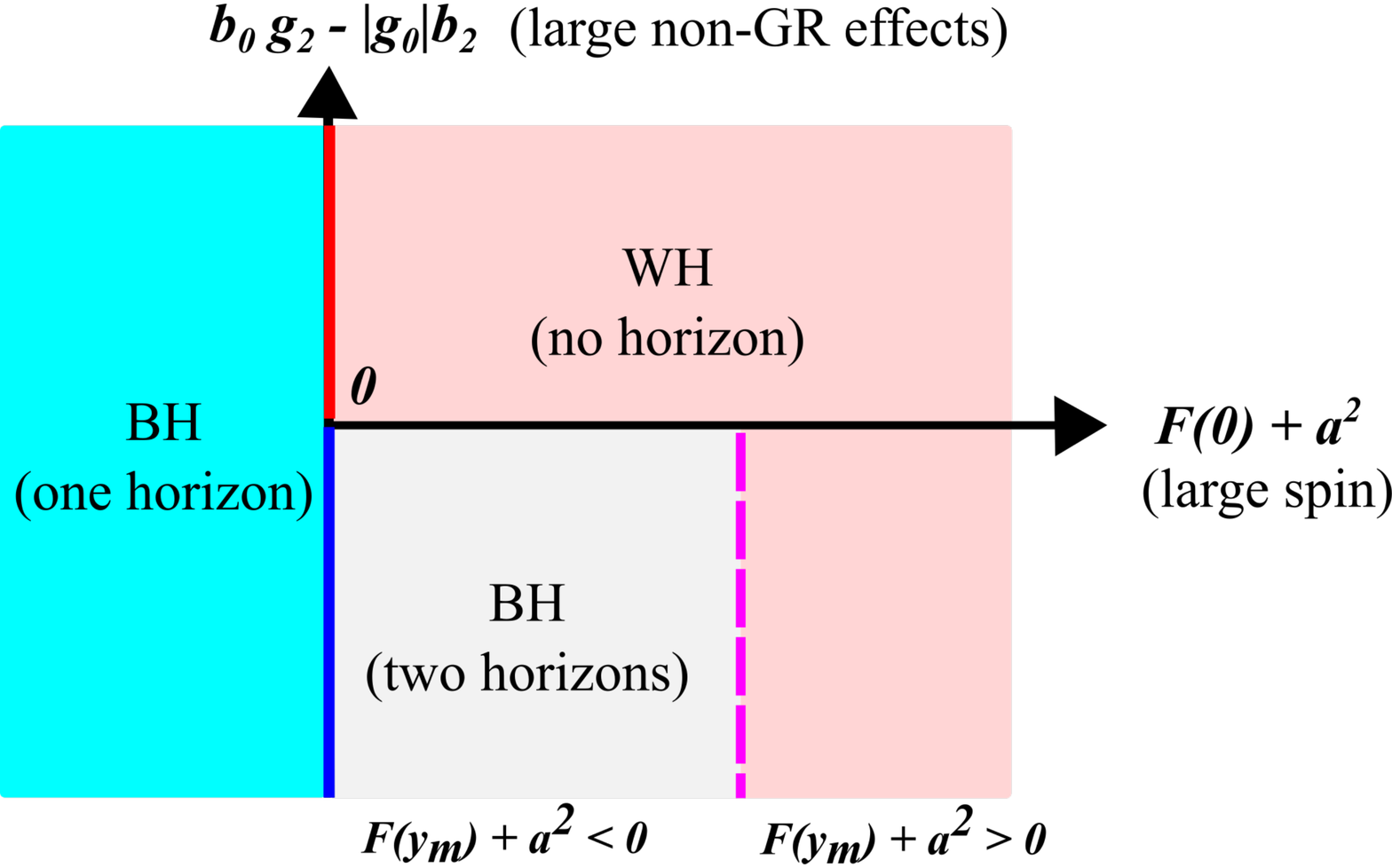,width=4in}}
\vspace*{8pt}
\caption{This figure shows the spacetime structure of the metric \eqref{NJAmetricfinal} within different regions of the parameter space.\label{fig1}}
\end{figure}

\begin{figure}[t]
\centerline{\psfig{file=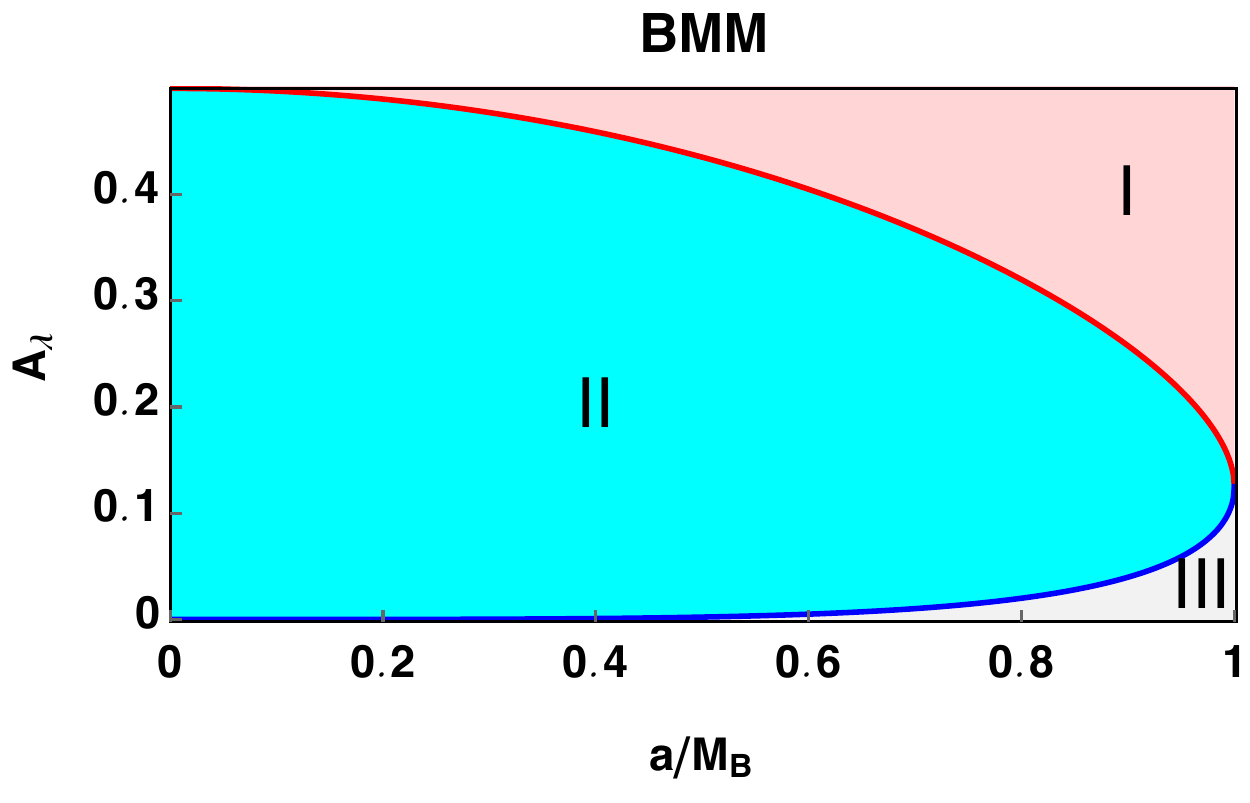,width=2.8in}\psfig{file=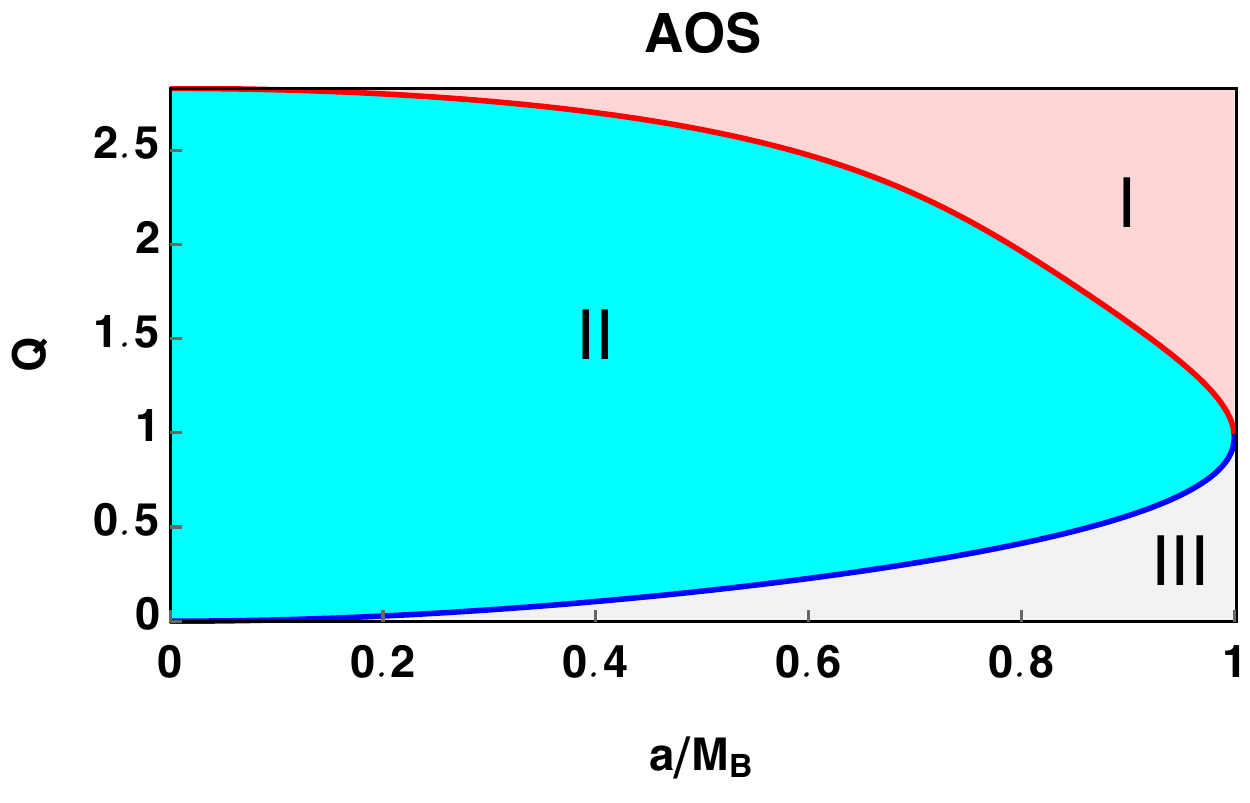,width=2.8in}}
\vspace*{8pt}
\caption{The spacetime structures of the rotating counterparts of the BMM (left) and the AOS (right) models, respectively. The rotating metrics are obtained through NJA. The vertical axis in each figure represents the quantum parameters of the models. The horizontal axis labels the spin of the black hole. Regions I, II, and III represent wormholes, black holes with one horizon, and black holes with two horizons, respectively (see Fig.~\ref{fig1}). \label{fig2}}
\end{figure}

The above discussions are summarized graphically in Fig.~\ref{fig1}. In this figure, the horizontal and the vertical axes denote the values of $F(0)+a^2$ and $b_0g_2-|g_0|b_2$, respectively. The parameter space toward the right of the horizontal axis corresponds to an increasing black hole spin. On the other hand, the parameter space toward the top of the vertical axis corresponds to larger deviations from Kerr metric, i.e., larger quantum parameters. This can be understood naively by the fact that the minimum value of the 2-sphere radius $b_0$ is expected to approach zero in the GR limit. In addition, the metric function $g(y)$ should approach to minus infinity in the same limit, i.e., $|g_0|\rightarrow\infty$ in the GR limit.

According to Fig.~\ref{fig1}, one can see that the values $F(0)+a^2$ and $b_0g_2-|g_0|b_2$ are zero at the intersection of the two axes. On the left of the vertical axis (cyan region), $F(0)+a^2$ is negative and the spacetime represents a black hole with only one horizon, behind which lies a spacelike transition surface. This includes the parameter space of non-rotating LQG black holes ($a=0$). 

When the value $F(0)+a^2$ is positive, as what we have mentioned before, depending on the sign of the value of $b_0g_2-|g_0|b_2$, the spacetime can represent a wormhole ($b_0g_2-|g_0|b_2>0$, pink region) or a black hole with two event horizons ($b_0g_2-|g_0|b_2<0$, light gray region). In particular, in the parameter space where $b_0g_2-|g_0|b_2<0$, if the spin is sufficiently large such that $F(y_m)+a^2$ is positive, the event horizon disappears and the spacetime represents a wormhole. Note that as long as $F(0)+a^2$ is positive, the transition surface is always timelike.

Furthermore, on the boundary between the cyan and the pink regions (red line), the transition surface itself is the only event horizon in the spacetime. On the other hand, on the boundary between the cyan and the light gray regions (blue line), there are two horizons, with the inner one being the transition surface itself. 

Before closing this section in which we have provided a general discussion on the possible spacetime structures for rotating LQG black holes, we would like to give two specific examples to support our arguments. We consider the non-rotating LQG black hole models proposed by Bodendorfer-Mele-M\"unch (BMM) \cite{Bodendorfer:2019nvy,Bodendorfer:2019jay} and Ashtekar-Olmedo-Singh (AOS) \cite{Ashtekar:2018lag,Ashtekar:2018cay,Ashtekar:2020ckv,Devi:2021ctm}. In both of these non-rotating black hole models, the classical Schwarzschild singularity is replaced by a spacelike transition surface inside the event horizon. We then construct their rotating counterparts using NJA and show their spacetime structures in Fig~\ref{fig2}. The horizontal axis in each figure represents the spin of the black hole. As for the vertical axes, on the other hand, the quantities $A_\lambda$ and $Q$ stand for the quantum parameters in the BMM model and the AOS model, respectively. One can clearly identify the cyan, pink, and light gray regions in each figure with their counterparts in Fig.~\ref{fig1}. The fine-tuned cases given by the red and blue curves can also be identified. This means that these two particular examples fit perfectly well the general discussion above. We expect that our general argument also applies to other LQG black hole models, as long as their non-rotating models have a transition surface inside the event horizon. Note also that similar spacetime structures can be obtained in a purely phenomenological way \cite{Mazza:2021rgq}, without resorting to quantum gravitational approaches.

\section{Conclusions}\label{sec.conclusion}
In this work, we give a conjecture on what the spacetime structure of a rotating LQG black hole may look like, based on a few general assumptions on metric functions of their non-rotating counterparts. The first assumption, and perhaps the most controversial one, is that the rotating metric can be constructed effectively by using NJA applied to a non-rotating seed metric. This assumption allows us, at least qualitatively, to address the physical consequences coming from rotating LQG black holes, given that a self-consistent LQG treatment on axisymmetric spacetimes is still lacking. Other assumptions that we have made include the assumption that the classical Schwarzschild singularity is replaced by a spacelike transition surface for non-rotating LQG black holes. This is again a fair assumption because the transition surface is a quite common consequence in many LQG black hole models. Finally, we assume that the quantum effects only concentrate near the transition surface. This assumption allows us to fairly restrict the behaviors of the seed metric functions.

Our argument can be understood as follows: In addition to the original event horizon, a non-zero spin value of the black hole would generically generate one more event horizon in the spacetime. Depending on the relative location of the transition surface and the two horizons, the spacetime structures of a rotating LQG black hole can become either a wormhole with a timelike transition surface, a black hole with one horizon and one spacelike transition surface, or a black hole with two horizons and one timelike transition surface. Our results are expected to fit most of effective LQG black holes, and are also expected to give some hints in the future developments of a more mathematically consistent construction of rotating LQG black hole models.

\section*{Acknowledgments}

CYC is supported by Institute of Physics in Academia Sinica, Taiwan.



\begin{thebibliography}{0}



\bibitem{Gambini:2013ooa}
R.~Gambini and J.~Pullin,
``Loop quantization of the Schwarzschild black hole,''
Phys. Rev. Lett. \textbf{110} (2013) no.21, 211301.

\bibitem{Corichi:2015xia}
A.~Corichi and P.~Singh,
``Loop quantization of the Schwarzschild interior revisited,''
Class. Quant. Grav. \textbf{33} (2016) no.5, 055006.

\bibitem{Olmedo:2017lvt}
J.~Olmedo, S.~Saini and P.~Singh,
``From black holes to white holes: a quantum gravitational, symmetric bounce,''
Class. Quant. Grav. \textbf{34} (2017) no.22, 225011.

\bibitem{Ashtekar:2018lag}
A.~Ashtekar, J.~Olmedo and P.~Singh,
``Quantum Transfiguration of Kruskal Black Holes,''
Phys. Rev. Lett. \textbf{121} (2018) no.24, 241301.

\bibitem{Ashtekar:2018cay}
A.~Ashtekar, J.~Olmedo and P.~Singh,
``Quantum extension of the Kruskal spacetime,''
Phys. Rev. D \textbf{98} (2018) no.12, 126003.

\bibitem{Bodendorfer:2019xbp}
N.~Bodendorfer, F.~M.~Mele and J.~M\"unch,
``A note on the Hamiltonian as a polymerisation parameter,''
Class. Quant. Grav. \textbf{36} (2019) no.18, 187001.

\bibitem{Bodendorfer:2019cyv}
N.~Bodendorfer, F.~M.~Mele and J.~M\"unch,
``Effective Quantum Extended Spacetime of Polymer Schwarzschild Black Hole,''
Class. Quant. Grav. \textbf{36} (2019) no.19, 195015.



\bibitem{Arruga:2019kyd}
D.~Arruga, J.~Ben Achour and K.~Noui,
``Deformed General Relativity and Quantum Black Holes Interior,''
Universe \textbf{6} (2020) no.3, 39.

\bibitem{Assanioussi:2019twp}
M.~Assanioussi, A.~Dapor and K.~Liegener,
``Perspectives on the dynamics in a loop quantum gravity effective description of black hole interiors,''
Phys. Rev. D \textbf{101} (2020) no.2, 026002.

\bibitem{BenAchour:2020gon}
J.~Ben Achour, S.~Brahma, S.~Mukohyama and J.~P.~Uzan,
``Towards consistent black-to-white hole bounces from matter collapse,''
JCAP \textbf{09} (2020), 020.

\bibitem{Gambini:2020nsf}
R.~Gambini, J.~Olmedo and J.~Pullin,
``Spherically symmetric loop quantum gravity: analysis of improved dynamics,''
Class. Quant. Grav. \textbf{37} (2020) no.20, 205012.


\bibitem{Bodendorfer:2019nvy}
N.~Bodendorfer, F.~M.~Mele and J.~M\"unch,
``(b,v)-type variables for black to white hole transitions in effective loop quantum gravity,''
Phys. Lett. B \textbf{819} (2021), 136390.

\bibitem{Bodendorfer:2019jay}
N.~Bodendorfer, F.~M.~Mele and J.~M\"unch,
``Mass and Horizon Dirac Observables in Effective Models of Quantum Black-to-White Hole Transition,''
Class. Quant. Grav. \textbf{38} (2021) no.9, 095002.

\bibitem{Blanchette:2020kkk}
K.~Blanchette, S.~Das, S.~Hergott and S.~Rastgoo,
``Black hole singularity resolution via the modified Raychaudhuri equation in loop quantum gravity,''
Phys. Rev. D \textbf{103} (2021) no.8, 084038.

\bibitem{Assanioussi:2020ezr}
M.~Assanioussi and L.~Mickel,
``Loop effective model for Schwarzschild black hole interior: a modified $\bar \mu$ dynamics,''
Phys. Rev. D \textbf{103} (2021) no.12, 124008.

\bibitem{Bojowald:2020dkb}
M.~Bojowald,
``Black-Hole Models in Loop Quantum Gravity,''
Universe \textbf{6} (2020) no.8, 125.


\bibitem{Boehmer:2007ket}
C.~G.~Boehmer and K.~Vandersloot,
``Loop Quantum Dynamics of the Schwarzschild Interior,''
Phys. Rev. D \textbf{76} (2007), 104030.

\bibitem{Bojowald:2018xxu}
M.~Bojowald, S.~Brahma and D.~h.~Yeom,
``Effective line elements and black-hole models in canonical loop quantum gravity,''
Phys. Rev. D \textbf{98} (2018) no.4, 046015.

\bibitem{BenAchour:2018khr}
J.~Ben Achour, F.~Lamy, H.~Liu and K.~Noui,
``Polymer Schwarzschild black hole: An effective metric,''
EPL \textbf{123} (2018) no.2, 20006.

\bibitem{Kelly:2020uwj}
J.~G.~Kelly, R.~Santacruz and E.~Wilson-Ewing,
``Effective loop quantum gravity framework for vacuum spherically symmetric spacetimes,''
Phys. Rev. D \textbf{102} (2020) no.10, 106024.


\bibitem{Frodden:2012en}
E.~Frodden, A.~Perez, D.~Pranzetti and C.~R\"oken,
``Modelling black holes with angular momentum in loop quantum gravity,''
Gen. Rel. Grav. \textbf{46} (2014) no.12, 1828.

\bibitem{Gambini:2020fnd}
R.~Gambini, E.~Mato and J.~Pullin,
``Axisymmetric gravity in real Ashtekar variables: the quantum theory,''
Class. Quant. Grav. \textbf{37} (2020) no.11, 115010.


\bibitem{Caravelli:2010ff}
F.~Caravelli and L.~Modesto,
``Spinning Loop Black Holes,''
Class. Quant. Grav. \textbf{27} (2010), 245022.

\bibitem{DeLorenzo:2015taa}
T.~De Lorenzo, A.~Giusti and S.~Speziale,
``Non-singular rotating black hole with a time delay in the center,''
Gen. Rel. Grav. \textbf{48} (2016) no.3, 31
[erratum: Gen. Rel. Grav. \textbf{48} (2016) no.8, 111].

\bibitem{Liu:2020ola}
C.~Liu, T.~Zhu, Q.~Wu, K.~Jusufi, M.~Jamil, M.~Azreg-A\"\i{}nou and A.~Wang,
``Shadow and quasinormal modes of a rotating loop quantum black hole,''
Phys. Rev. D \textbf{101} (2020) no.8, 084001
[erratum: Phys. Rev. D \textbf{103} (2021) no.8, 089902].

\bibitem{Brahma:2020eos}
S.~Brahma, C.~Y.~Chen and D.~h.~Yeom,
``Testing Loop Quantum Gravity from Observational Consequences of Nonsingular Rotating Black Holes,''
Phys. Rev. Lett. \textbf{126} (2021) no.18, 181301.





\bibitem{Newman:1965tw}
E.~T.~Newman and A.~I.~Janis,
``Note on the Kerr spinning particle metric,''
J. Math. Phys. \textbf{6} (1965), 915-917.

\bibitem{Azreg-Ainou:2014pra}
M.~Azreg-A\"\i{}nou,
``Generating rotating regular black hole solutions without complexification,''
Phys. Rev. D \textbf{90} (2014) no.6, 064041.



\bibitem{Yang:2021civ}
R.~Q.~Yang, R.~G.~Cai and L.~Li,
``Constraining the number of horizons with energy conditions,''
[arXiv:2104.03012 [gr-qc]].


\bibitem{Ashtekar:2020ckv}
A.~Ashtekar and J.~Olmedo,
``Properties of a recent quantum extension of the Kruskal geometry,''
Int. J. Mod. Phys. D \textbf{29} (2020) no.10, 2050076.


\bibitem{Devi:2021ctm}
S.~Devi, A.~N.~S, S.~Chakrabarti and B.~R.~Majhi,
``Shadow of quantum extended Kruskal black hole and its super-radiance property,''
[arXiv:2105.11847 [gr-qc]].


\bibitem{Mazza:2021rgq}
J.~Mazza, E.~Franzin and S.~Liberati,
``A novel family of rotating black hole mimickers,''
JCAP \textbf{04} (2021), 082.

\end{thebibliography}
\end{document}